\documentclass[10pt, conference, compsocconf]{IEEEtran}
\usepackage{mathptmx} 

\usepackage{fancyhdr}
\usepackage[normalem]{ulem}
\usepackage[hyphens]{url}
\usepackage[sort,nocompress]{cite}
\usepackage[final]{microtype}
\usepackage[keeplastbox]{flushend}

\usepackage{colortbl}
\usepackage{braket}
\usepackage{arydshln}
\usepackage{float}
\usepackage{courier}
\usepackage{comment}
\usepackage{amsmath,amssymb,amsfonts}
\usepackage{algorithmic}
\usepackage{braket}
\usepackage{graphicx}
\usepackage{textcomp}
\usepackage[dvipsnames]{xcolor}
\usepackage{multirow}
\usepackage{multicol}
\usepackage{soul}

\usepackage{tikz}
\newcommand*\circled[1]{\tikz[baseline=(char.base)]{
            \node[shape=circle,draw,inner sep=0.5pt] (char) {#1};}}

\usepackage{booktabs}

\makeatletter 
\newcommand{\linebreakand}{%
  \end{@IEEEauthorhalign}
  \hfill\mbox{}\par
  \mbox{}\hfill\begin{@IEEEauthorhalign}
}
\makeatother 

\newcommand{\subparagraph}{}

\usepackage[bookmarks=true,breaklinks=true,letterpaper=true,colorlinks,citecolor=blue,linkcolor=blue,urlcolor=blue]{hyperref}

\ifCLASSINFOpdf
\else
\fi
\begin{document}
%
\title{Scaling Superconducting Quantum Computers with Chiplet Architectures}


\author{\IEEEauthorblockN{Kaitlin N. Smith}
\IEEEauthorblockA{
\normalfont{University of Chicago}\\
Chicago, USA\\
kns@uchicago.edu}
\and
\IEEEauthorblockN{Gokul Subramanian Ravi }
\IEEEauthorblockA{
\normalfont{University of Chicago}\\
Chicago, USA\\
gravi@uchicago.edu}
\and
\IEEEauthorblockN{Jonathan M. Baker}
\IEEEauthorblockA{
\normalfont{University of Chicago}\\
Chicago, USA\\
jmbaker@uchicago.edu}
\and
\IEEEauthorblockN{Frederic T. Chong}
\IEEEauthorblockA{
\normalfont{University of Chicago}\\
Chicago, USA\\
chong@cs.uchicago.edu}
}


%


\maketitle

\begin{abstract}
Fixed-frequency transmon quantum computers (QCs) 
have advanced in coherence times, addressability, and gate fidelities. Unfortunately, these devices are restricted by the number of on-chip qubits, capping processing power and slowing progress toward fault-tolerance. Although emerging transmon devices feature over 100 qubits, building QCs large enough for meaningful demonstrations of quantum advantage requires overcoming many design challenges.
For example, today's transmon qubits suffer from significant variation due to limited precision in fabrication. As a result, 
barring significant improvements in current fabrication techniques, scaling QCs by building ever larger individual chips with more qubits is hampered by device variation.
Severe device variation that degrades QC performance is referred to as a defect. Here, we focus on a specific defect known as a frequency collision. 

When transmon frequencies collide, their difference falls within a range that limits two-qubit gate fidelity. 
Frequency collisions occur with greater probability on larger QCs, causing collision-free yields to decline as the number of on-chip qubits increases. As a solution, we propose exploiting the higher yields associated with smaller QCs by integrating quantum chiplets within quantum multi-chip modules (MCMs). Yield, gate performance, and application-based analysis show the feasibility of QC scaling through modularity. Our results demonstrate that chiplet architectures, relative to monolithic designs, benefit from average yield improvements ranging from  $9.6-92.6\times$ for $\lesssim$500 qubit machines. In addition, our simulations explore the design space of chiplet systems and discover configurations that demonstrate average two-qubit gate infidelity reductions that are at best $0.815\times$ their monolithic counterpart. Finally, we observe that carefully-selected modular systems achieve fidelity improvements on a range of benchmark circuits.

\end{abstract}

\begin{IEEEkeywords}
Quantum computing; quantum architecture; superconducting quantum computers

\end{IEEEkeywords}

%
\IEEEpeerreviewmaketitle

\section{Introduction}
\label{1-introduction}

The proposed benefits of quantum information processing has garnered much interest from industry, academia, and government alike. Specifically, quantum computers (QCs) 
are forecast to accelerate the solving of some of today's most 
complex problems through the careful use of quantum superposition, interference, and entanglement. Fault-tolerant QCs have been proposed for applications in cryptography~\cite{shor1999polynomial}, big data~\cite{grover1996fast} chemistry~\cite{kandala2017hardware}, optimization~\cite{moll2018quantum}, and machine learning~\cite{biamonte2017quantum}. Although current QCs are noisy prototypes without error correction, there exist early, yet exciting, demonstrations of quantum advantage~\cite{arute2019quantum,zhong2020quantum}.

The distinction between classical and quantum processors will become more apparent as scaled QCs emerge that can access exponentially larger computational spaces. 
While QCs with over 100 physical qubits have been built~\cite{eagle-127q,eagle-processor}, 
 far more qubits are required to solve meaningful, real-world problems at rates that outperform or even compete with existing classical hardware. Additionally, near-term QCs, referred to as noisy intermediate scale quantum (NISQ) machines~\cite{preskill2018quantum}, have high gate infidelity and low coherence times that prohibit error correction.
Large-scale quantum algorithms such as Shor's Factoring~\cite{Shor_1997} and Grover's Search~\cite{Grover96afast} will require millions of fault-tolerant qubits for computation~\cite{O_Gorman_2017}.

\begin{figure}[t]
     \centering
         \includegraphics[width=0.97\columnwidth,clip]{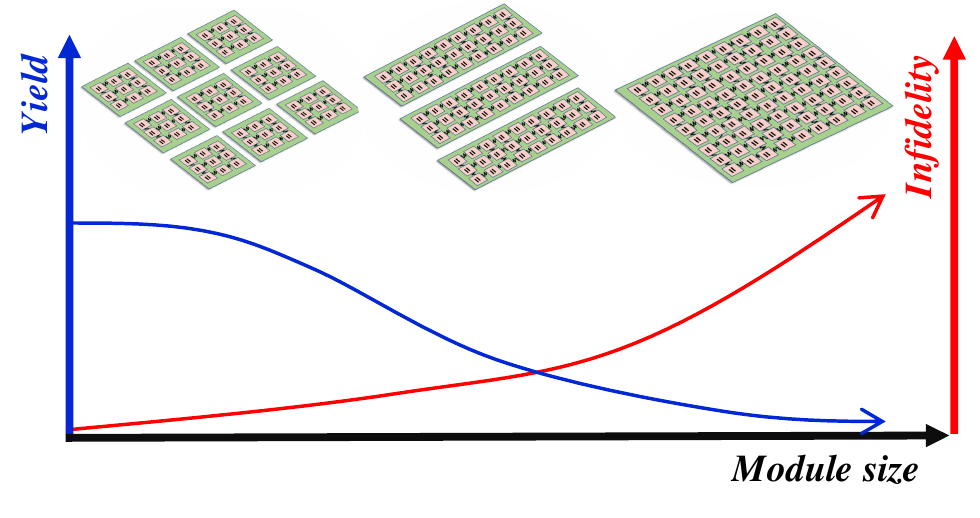}
        \caption{Infidelity and yield trade-off vs. individual QC module size in qubits. 
        }
        \label{fig:infidelity-vs-size} 

\end{figure}

Scaling QCs requires confronting many challenges including, but not limited to, 
qubit quality, external noise, power consumption, control routines, device packaging, hardware footprint, and system cooling~\cite{national2019quantum}. In the near term, especially on QCs based on superconducting (SC) circuits, limitations associated with qubit fabrication are among the most debilitating on the path to QC scaling. 
The physical processing that attempts to realize targeted, per-qubit specifications suffers from 
 unavoidable variation. This fabrication imprecision 
 can inject random defects into QCs~\cite{hertzberg2021laser,devoret2004superconducting}, severely impacting QC fidelity. As fabricated chips increase in size, imprecision-related defects become statistically more likely. 
Thus, intuition expects that average device infidelity correlates with device size, as illustrated on the right axis in Fig.~\ref{fig:infidelity-vs-size}. Eventually, infidelity surpasses usefulness thresholds for a QC, causing fabrication yields to decline,
 as shown on the left of Fig.~\ref{fig:infidelity-vs-size}.

Fixed-frequency transmon QCs, like those in the IBM Quantum systems~\cite{IBMQS}, are considered promising for scaling. Transmons are constructed of superconducting (SC) circuits, and they take advantage of fabrication materials and processes refined 
for classical integrated circuits. Transmons also have fairly good coherence times that reliably extend to tens of microseconds~\cite{houck2008controlling,chang2013improved,IBMQS} along with gate fidelities that are approaching fault-tolerance thresholds~\cite{sheldon2016procedure}. Although they have many benefits, transmons are not exempt from fabrication-related scaling challenges. 
In this work we primarily focus on frequency collisions.

Frequency collisions are a specific type of defect commonly seen in fixed-frequency transmon QCs. A frequency collision is caused by a qubit-qubit frequency detuning that falls in a range that degrades two-qubit gate fidelity~\cite{brink2018device}.
Stocastic variation in QC fabrication influences the operational characteristics of Josephson Junctions (JJs), or the essential components of transmons that realize qubit state. Intrinsic variation alters both JJ dimension and critical current, and as a result, actual qubit frequency drifts from its intended value~\cite{kreikebaum2020improving}. This causes collision conditions that restrict QC operation because qubits lose their ability to participate in high-quality multi-qubit entangling operations.

Larger QCs with more qubits experience greater variation during fabrication. This variation accumulates and introduces error channels in the form of QC defects.
In the case of frequency collisions, the probability of a collision occurring on chip increases with QC size~\cite{hertzberg2021laser,zhang2020high}, and thus, the collision-free yield decreases rapidly. 
This insight motivates our work. As a solution, we propose techniques for transmon QC scaling that boost device yields and performance through modularity. 
At time of writing, scaling through modularity has gained traction within concurrently-released quantum roadmaps~\cite{roadmap-to-2025}, and
in our study, we provide a novel analysis of quantum chiplets and quantum multi-chip modules (MCMs) that considers the physical effects limiting yield and application fidelity. MCMs can be thought of in a similar manner as multi-core systems and 
are designed to exploit the larger yields of smaller quantum chiplets. 
In some extremes, our simulations show that MCM architectures can provide fidelity gains and construct QCs of dimensions that result in zero monolithic yield. 
Our contributions are as follows: 

\circled{1} We describe current fidelity trends of today's QCs based on real-machine data, showing that smaller QCs are favored in terms of minimized variation and gate error. 

\circled{2} We investigate prior work related to classical scaling via modular systems. We draw parallels to transmon QC technology and current challenges that exist with scaling. 
    
\circled{3} We identify frequency collision conditions 
and
critical parameters in state-of-art device fabrication. We explore the frequency allocation and fabrication precision design space, highlighting their influence on collision-free yield. 

\circled{4} Motivated by frequency collision defects, we propose a flexible chiplet design for integration within large-scale transmon QCs. We keep eventual fault-tolerance in mind with chip bonding that preserves heavy-hex connectivity. 

\circled{5} We introduce the customizable, quantum multi-chip module (MCM) and propose emerging link technology for chip unification. 

\circled{6} We develop a simulation framework that models monolithic and MCM QCs for yield, performance, and application-based analysis. This framework will be open sourced and is highly parameterized for adaptability: simulations and QCs can be programmed to model future improvements in fabrication and devices.

\circled{7} We provide simulation results that show significant collision-free yield improvements for chiplets and, thus, MCMs. 
Analysis using real quantum benchmarks and state-of-art error metrics showcase situations where MCM properties and performance surpass monolithic counterparts on top of yield advantages, showing points of crossover in terms of architecture of choice. 

The models presented here are extremely timely: modularity is gaining widespread acceptance as the best route to quantum scaling~\cite{roadmap-to-2025,rodrigo2020will}. In our study, we hope to guide the development of modular QCs, showing the feasibility of chiplet-based MCMs via yield, performance, and application-based analysis.  
Our results show that chiplet architectures, relative to monolithic designs, benefit from average yield improvements ranging from  $9.6-92.6\times$ for $\lesssim$500 qubit machines. In addition, our simulations explore the design space of chiplet systems and discover configurations that demonstrate average two-qubit gate infidelity reductions $ 0.949-0.815\times$ of their monolithic counterparts. Finally,
we observe that carefully-selected modular systems achieve
fidelity improvements on a range of benchmark circuits.

\section{Background}
\label{2-background}

\subsection{Quantum Information}
\label{information-qc}

The qubit can hold a superposition of the basis states $\ket{0}$ and $\ket{1}$ until measurement where the state $\ket{\psi} = \alpha\ket{0} + \beta\ket{1}$ collapses into 
either $\ket{0}$ or $\ket{1}$. 
$CX$ along with single-qubit rotation operations are frequently supported 
on near-term QCs, and
quantum circuits are cascades of these gates. 

Limitations of today's quantum hardware in terms of coherence and noise force a one-to-one mapping between logical qubits of an algorithm to physical qubits on a QC. A goal in quantum hardware design is to achieve error rates below the thresholds required for fault-tolerance via error correcting codes (ECCs). ECCs encode a logical qubit on many physical qubits, smoothing occasional gate and decoherence errors on individual physical qubits. Prior work estimates that at minimum, millions of high-quality physical qubits will be required for fault-tolerant (FT) quantum computation~\cite{o2017quantum}. Once FT QCs emerge, disruptive quantum applications in areas such as cryptography~\cite{shor1999polynomial} and big data~\cite{grover1996fast} will be feasible. More quantum computing fundamentals can be found in \cite{mike_ike_2020}.

\begin{table*}
\begin{center}
\resizebox{1.8\columnwidth}{!}{%
\begin{tabular}{ c | c | c | c | c }
\textbf{Type} & \textbf{Criteria} & \textbf{Threshold} & \textbf{Note} & \textbf{Qubit Topology}\\
\hline
1 & $f_i = f_j$ & $\pm 0.017$ GHz & Nearest neighbors $Q_i$ and $Q_j$& \multirow{8}{*}{\includegraphics[width=0.4\columnwidth,trim={0cm 0cm 0cm 0cm},clip]{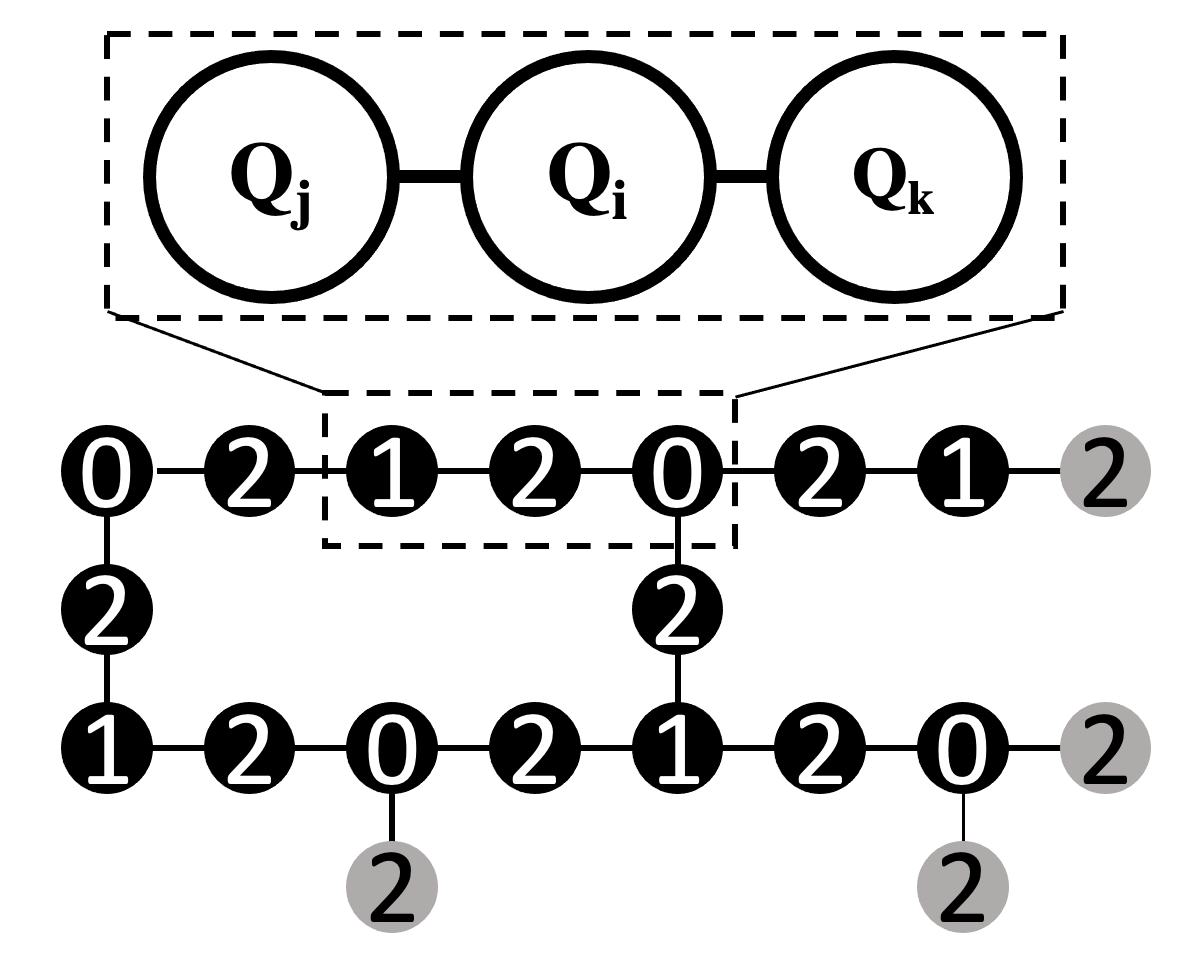}} \\

2 & $f_i +\alpha/2 = f_j$&$\pm 0.004$ GHz & Nearest neighbors control $Q_i$ and target $Q_j$   \\

3 & $f_i=f_j +\alpha$&$\pm 0.03$ GHz & Nearest neighbors $Q_i$ and $Q_j$ \\

\multirow{2}{*}{4} & $f_j<f_i+\alpha$& \multirow{2}{*}{-} & \multirow{2}{*}{Nearest neighbors control $Q_i$ and target $Q_j$}\\
 & \textbf{\emph{or}} $f_i< f_j$ & & \\

5 &$f_j=f_k$ & $\pm 0.017$ GHz & $Q_i$ is control to nearest neighbors $Q_j$ and/or $Q_k$  \\

\multirow{2}{*}{6} &$f_j=f_k+\alpha$ &\multirow{2}{*}{$\pm 0.025$ GHz} &\multirow{2}{*}{$Q_i$ is control to nearest neighbors $Q_j$ and/or $Q_k$}\\
 & \textbf{\textit{or}} $f_j+\alpha=f_k$ & & \\

7 &$2f_i +\alpha = f_j + f_k$ & $\pm 0.017$ GHz & $Q_i$ is control to nearest neighbors $Q_j$ and/or $Q_k$\\
\end{tabular}%
}
\end{center}
\caption{Fixed-frequency transmon collision criteria to bound gate error from frequency-related noise to $ \lesssim 1\%$. 
}
\label{tab:collision criteria}
\end{table*}

\subsection{Transmon Qubits}
\label{sect:transmon-qubits}

Fixed-frequency transmon devices 
provide an especially viable platform for realizing physical qubits.
 Compared to alternative technologies, transmon QCs show great promise due to recent breakthroughs in device coherence, operation fidelity, and addressability~\cite{mckay2017efficient,kjaergaard2020superconducting,jurcevic2021demonstration}.  
 Transmon devices are also favorable to scaling as they take advantage of fabrication techniques refined by 
 classical electronics. 
By employing Josephson Junctions (JJs) within SC circuits, transmons realize quantum information by acting as mesoscopic-scale, artificial atoms characterized by an anharmonic energy spectrum. Transmon qubits have many properties that must be properly characterized for successful control. 
 
The spectrum associated with a transmon qubit contains energy levels beyond the (lowest) two that are used for information encoding. 
 A transmon's operation frequency, $f_i$, is set by the the transition 
 between the qubit $\Ket{1}$ and $\ket{0}$ state, $f_i = f_{i,01} = f_{\ket{1}}-f_{\ket{0}}$. Operation frequency is determined by the anharmonic energy levels of the qubit. Transmon anharmonicity, $\alpha$ is fixed by the energy level difference between the $\ket{1}\rightarrow\ket{2}$ transition and the $\ket{0}\rightarrow\ket{1}$ transition, $\alpha = f_{i,12}-f_{i,01}$. Today's state-of-art transmon devices target $f_i$ values in the neighborhood of  $\sim 5$ GHz while qubit anharmonicity is $\alpha \approx -0.330$ GHz~\cite{zhang2020high}. 
 
 The IBM transmon QCs are up to 127 qubits in size~\cite{eagle-processor}. Microwave drives implement single-~\cite{chow2010optimized} and two-qubit gates~\cite{chow2011simple} on these devices with fidelities that are moving toward FT thresholds. 
 While 
 all-to-all physical qubit connectivity is desirable, design constraints associated with limited space on-chip  
 restricts near-term transmon qubits to communication with their nearest neighbors. It might be assumed that QC topologies would be designed for the highest qubit degree possible, but highly-connected qubits have an elevated risk of crosstalk, or unintended classical or quantum coupling, that degrades computation~\cite{chamberland2020topological}. As a solution, IBM QCs implement heavy-hexagon, or heavy-hex, topologies to enable sufficient physical qubit-qubit communication while minimizing cost in terms of crosstalk and hardware requirements~\cite{heavy-hex-blog}. An example of three IBM QCs that use the heavy-hex qubit layout can be found in Fig.~\ref{fig:generation-comparison}(a). The heavy-hexagon lattice is targeted for the hybrid surface/Bacon-Shor ECC with an error threshold of 0.45\%~\cite{chamberland2020topological}. 

\begin{figure}[b]
     \centering
         \includegraphics[width=0.97\columnwidth,trim={0cm 0cm 0cm 0cm},clip]{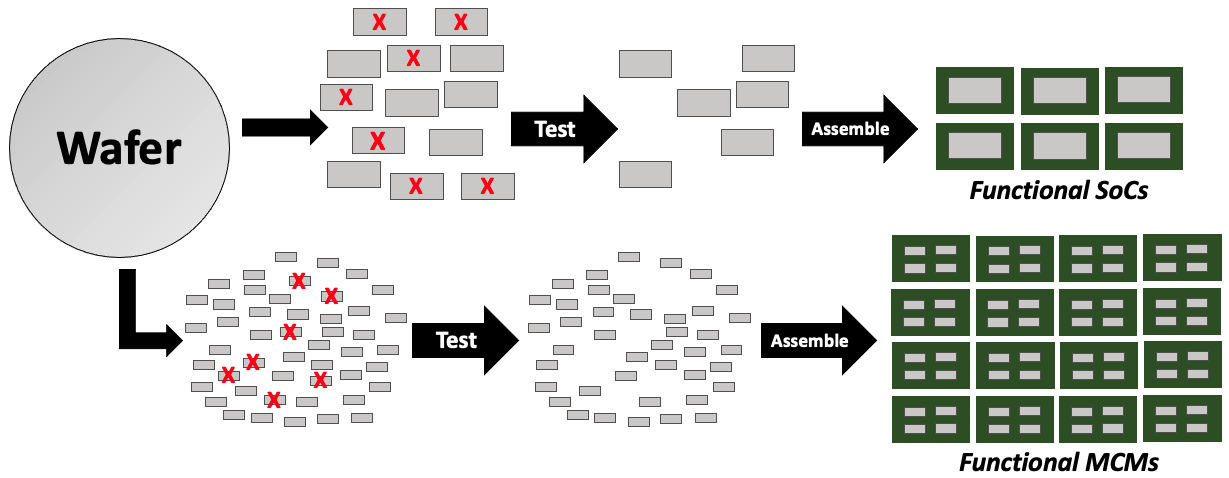}
        \caption{Monolithic (top) vs. chiplet (bottom) architecture yields with seven faulty devices in each batch.}
        \label{fig:chiplet-vs-monolithic}
\end{figure}

\subsection{Modularity in Classical Scaling}

Classical computing systems have drastically improved since their emergence in the mid 1900s. Thus, we are motivated to examine classical methodologies to find quantum parallels that could lead to accelerated QC scaling.

Monolithic electronics have benefits that stem from the simplicity of being entirely self-contained as a system on a chip (SoC). SoCs, however, have added costs associated with yields~\cite{leachman2011yield}. Additionally, monolithic SoCs often have steep resource requirements~\cite{esmaeilzadeh2011dark} and demonstrate inflexibility in implementing highly customized or specialized hardware as the SoC footprint increases~\cite{chiplet-blog}. Fortunately, chiplet-based architectures can alleviate some of these problems, pushing forward architectural innovations that allow new computing milestones~\cite{demir2014galaxy, chiplet-blog, kim2007architectural,pal2021designing,yin2018modular}. In a chiplet design, multiple smaller and locally-connected chips replace a larger, monolithic SoC. Chiplet architectures are often referred to as multi-chip modules (MCMs).

A few benefits of chiplet are as follows. First, chiplet architectures take advantage of well-established electronic manufacturing techniques because each chiplet, like its monolithic counterpart, is an integrated circuit (IC). Second, as seen in Fig.~\ref{fig:chiplet-vs-monolithic}, the smaller-footprint chiplet has 
lower fabrication cost that results in a higher yield of functional devices since significantly more can be fabricated simultaneously on a substrate. Third, because chips contain fewer components, the complexity of design verification and post-fabrication test is reduced. Fourth, chiplets enable mixed-process integration  
within a single package, potentially allowing significant processing gains. Because individual chiplets contain less logic, application-specific components can utilize more optimized technology libraries and acceleration that would not provide advantage to more generic logic. Finally, since chiplet designs are modular, failures can be isolated and removed without the need to discard the system as a whole.  

Chiplet-based classical computation offers many advantages that promote scaling. However, communication bottlenecks, often in the form of latency in link hardware, must be properly evaluated as a trade-off. MCMs sometimes demonstrate performance loss as compared to their monolithic counterparts~\cite{kim2007architectural}, but the benefits associated with modularity are often overwhelmingly worth the cost.

\section{Transmon QC Challenges}
\label{3-mot-scaling-transmon}

This section focuses on three interconnected transmon constraints that currently impede progress toward realizing larger transmon QCs: Cross-Resonance gate error, architecture rigidity, and fabrication imprecision.

\subsection{The Cross-Resonance Gate}
\label{sect:CR-gate}

A hallmark of a promising quantum implementation is strong qubit-qubit entanglement. Transmons can become entangled after executing the Cross-Resonance ($CR$) gate on a pair of qubits connected by a resonator~\cite{paraoanu2006microwave,rigetti2010fully,magesan2020effective,tripathi2019operation}. This gate operates by applying a microwave drive to one qubit, the control, at the frequency of a second, the target.  
The ideal $CR$ effect is a $ZX$ interaction between the participating qubits. This interaction can then be combined with single-qubit operations to achieve more standard two-qubit gates used in circuits, such as the $CX$ gate. 

$CR$ error rates lower than $10^{-2}$ have been reported~\cite{sheldon2016procedure,kandala2021demonstration,jurcevic2021demonstration}, but $CR$ gate implementation must improve to reach FT QCs. $CR$ error rates can theoretically reach levels as low as $10^{-4}$~\cite{malekakhlagh2020first}. Reaching this lower bound is necessary to improve QC scaling prospects. However, 
$CR$ fidelity is limited if system noise is inaccurately characterized, leading to coarse models used during $CR$ optimization. Despite advances made in qubit optimal control at the pulse-level, unexpected environmental coupling or irregular QC hardware properties stemming from manufacturing inconsistency or defects can significantly influence $CR$ error.

\subsection{Architectural Constraints}
\label{sec:arch-constraints}

Fixed-frequency transmon devices must be carefully designed in terms of their frequency assignment to 1) allow for selective addressability among qubits and 2) to enable high-quality $CR$ interactions between nearest neighbors. Table~\ref{tab:collision criteria} lists seven conditions that include numerically defined definitions and bounds from~\cite{hertzberg2021laser,magesan2020effective} that are likely to lead to $\gtrsim1\%$ gate error due to transmon qubit \emph{frequency collisions}, or conditions where qubit frequencies ``collide,'' degrading $CR$ performance. Frequency collisions are a leading challenge in scaling transmon devices~\cite{zhang2020high}. 
As a note, current transmon systems on-average demonstrate $\sim$1-2\% infidelity contributed by various noise sources unrelated to error caused by frequency~\cite{takita2017experimental,mckay2019three}. 

The criteria of Table~\ref{tab:collision criteria} define fabrication targets that minimize error stemming from $CR$ interactions between qubits with improper frequency spacing. Violation of Table~\ref{tab:collision criteria} criteria amplifies gate infidelity considerably as the frequency collision becomes the dominant source of $CR$ error. Careful frequency assignment in terms of targeted individual frequency and nearest neighbor detuning during the QC design process helps prevent collisions. Unfortunately, current limitations in fabrication precision inject random variation, causing actual QCs to have features that deviate from what is theoretically expected. More information about fabrication variation can be found in Section~\ref{sec:device-fab-var}. 

\textbf{Ideal Frequency Assignment:} To avoid collisions, each qubit's frequency must be distinguishable from its neighbor. It is also desirable to minimize the number of target frequencies, $F_i$, on chip to simplify control. Frequency collision conditions can be avoided using three frequencies in the heavy-hex lattice where $F_0<F_1<F_2$~\cite{hertzberg2021laser}. Here, the highest frequency qubits, $F_2$, act as control qubits in $CR$ interactions and never have a degree greater than two. As a note, unidirectional two-qubit interactions are natively supported by hardware, but the direction of two-qubit operations can be reversed with single-qubit rotation operations. 

The ideal heavy-hex frequency assignment is pictured in the lower half of Table~\ref{tab:collision criteria}, column 5. As the pictured QC topology scales, the frequency assignment pattern remains constant. To avoid collisions, nearest neighbor frequencies should never be equivalent, and all $F_2$ qubits should be surrounded by $F_0$ and $F_1$ but never by two of the same kind.

\textbf{Frequency Detuning:} The frequency difference of qubits must fall within a ``straddling regime'' to avoid collisions: too small of a frequency detuning results in inaddressability and too far prevents entanglement from being created~\cite{magesan2020effective}. Second, the relationship between the frequencies of next-nearest neighbors is also of importance. For example, consider the three qubits in the top half of Table~\ref{tab:collision criteria}, column 5. If a control transmon, $f_i=F_2$, is connected to two qubits of almost equivalent frequencies, $f_j=F_1 \approx f_k=F_0$, applying a drive of $f_j$ to the control would cause a frequency collision since a high-noise $CR$ interaction results if $f_j$ is near-resonant to $f_k$. This is an example of a ``near-null,'' frequency collision, or a Type 1 or 5 collision in Table~\ref{tab:collision criteria}.

\begin{figure*}[!th]
     \centering
         \includegraphics[width=0.97\textwidth,trim={0.12cm 6.9cm 0.12cm 6.8cm},clip]{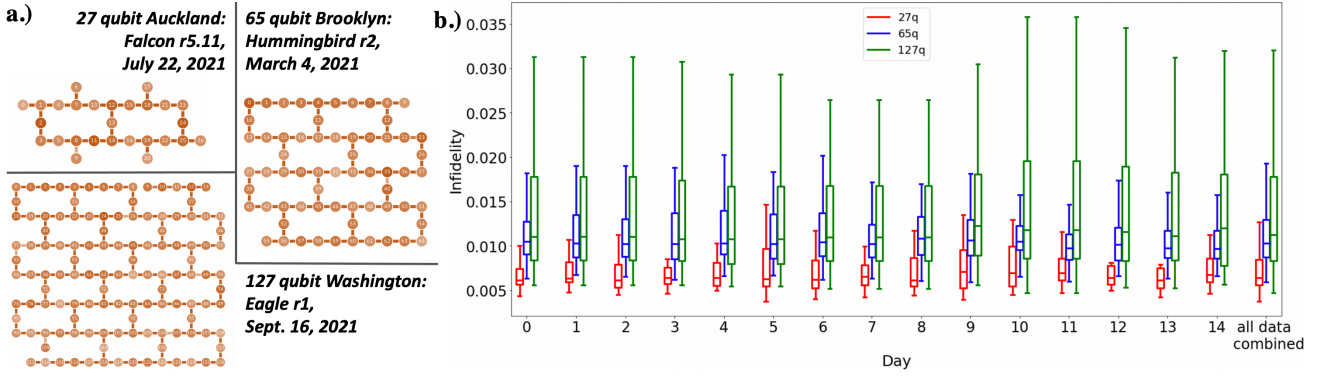}
        \caption{(a) Device name, processor type, date online, and connectivity graph for three IBM QCs. (b) Box plot comparing of 15 days of CX infidelity statistics for three generations of IBM processors of different sizes.}
        \label{fig:generation-comparison}

\end{figure*}

\subsection{Device Fabrication Variation}
\label{sec:device-fab-var}

Precise control over qubit frequency is critical to produce high-fidelity transmon QCs. In reality, this can be extremely challenging due to the small feature size, approximately 100$\times$100 nm, of the transmon JJ~\cite{hertzberg2021laser}. Small imperfections appear in JJ positioning and 
dimensions during processing, influencing the frequency at which the fixed-frequency transmon operates~\cite{kreikebaum2020improving}. Fabrication variation that prohibits or severely limits a transmon qubit's functionality is known as a defect. Here, we will focus on defects that deviate a qubit's frequency from its ideal, resulting in spectral overlaps that cause frequency collisions. This variation is stocastic, causing the final frequency profile of each chip to be unique. The distribution that describes fabrication precision is characterized by the standard deviation $\sigma_f$, where a lower $\sigma_f$ indicates higher precision characterized by actual device frequencies landing near design targets. Shortcomings in fabrication accuracy currently result in devices with a frequency spread of $\sim 0.1$ GHz around targeted frequencies~\cite{zhang2020high}. This large amount of variation causes qubit-qubit frequency detunings that satisfy the Table~\ref{tab:collision criteria} criteria and drive down the number of functional QC devices in the collision-free yield. With such high variation, yields reach zero well before devices reach $10^2$ qubits. For example, a real-world fabrication variation of $\sigma_f= 0.1323$ GHz~\cite{hertzberg2021laser} is seen as the blue curve in the plots of Fig.~\ref{fig:detuning-compare}. At this poor precision, there is little hope of creating high-yield quantum chips containing more than 20 qubits. Fortunately, transmon qubit frequency can be characterized and laser tuned after fabrication, enabling a much higher level of precision characterized by $\sigma_f\approx 0.014$ GHz, boosting yield~\cite{hertzberg2021laser}. Laser tuning has been applied to improve $<100$ qubit QC yields by 15$\times$~\cite{zhang2020high}. Further improvements to $\sigma_f$ 
are required to maintain yield while increasing the number of qubits manufactured on a monolithic QC.

\section{Motivating Quantum Chiplets}
\label{motivate-chiplets}
\subsection{Recent Trends in Quantum Hardware}
 
Calibration data for three IBM machines of different processor generations was gathered to analyze relationships between $CX$ gate performance and processor size. The machines included Auckland, a 27-qubit Falcon, Brooklyn, a 65-qubit Hummingbird, and Washington, a 127 qubit Eagle. Details of these machines, including date online, revision, and heavy-hexagon connectivity graphs, are featured in Fig.~\ref{fig:generation-comparison}(a).  All QCs under analysis were released in 2021 and their size correlates to processor generation: larger devices with more qubits feature newer hardware designs~\cite{IBM-processor-types}. Statistics detailing $CX$ infidelity for each device over the course of 15 days, i.e. 15 different calibration cycles, are found in Fig.~\ref{fig:generation-comparison}(b). 

Newer IBM processors feature technological advancements for improved scaling. For example, hardware upgrades enable computational benefits such as longer qubit coherence, reduced gate and readout times, and smaller footprint with innovations such as ``direct couplers''~\cite{IBM-processor-types,eagle-processor}. However, Fig.~\ref{fig:generation-comparison}(b) shows that median $CX$ infidelity directly correlates with chip size such that larger devices demonstrate higher two-qubit gate error rates over the 15 calibration cycles. Moreover, the $CX$ infidelity of the larger devices generally exhibits a larger distribution of $CX$ quality over the investigation period. These results suggest that greater variation exists as QCs increase in size, degrading overall performance. As described in Section~\ref{3-mot-scaling-transmon}, fabrication variation exists in near-term QCs, and the extent of heterogeneity of qubits becomes more extreme as devices increase in size. We postulate that as fabrication precision stays constant, larger devices inadvertently have the potential to contain more qubit-qubit frequency detunings that push them within or close to frequency collision regions. Thus, we are motivated to pursue greater consistency and performance in machines through modularity.

\subsection{Modeling Yield vs. Qubits}
\label{yield-vs-qubits}

\begin{figure*}[t!]
     \centering
         \includegraphics[width=0.97\textwidth,trim={0cm 4.7cm 0.05cm 4.7cm},clip]{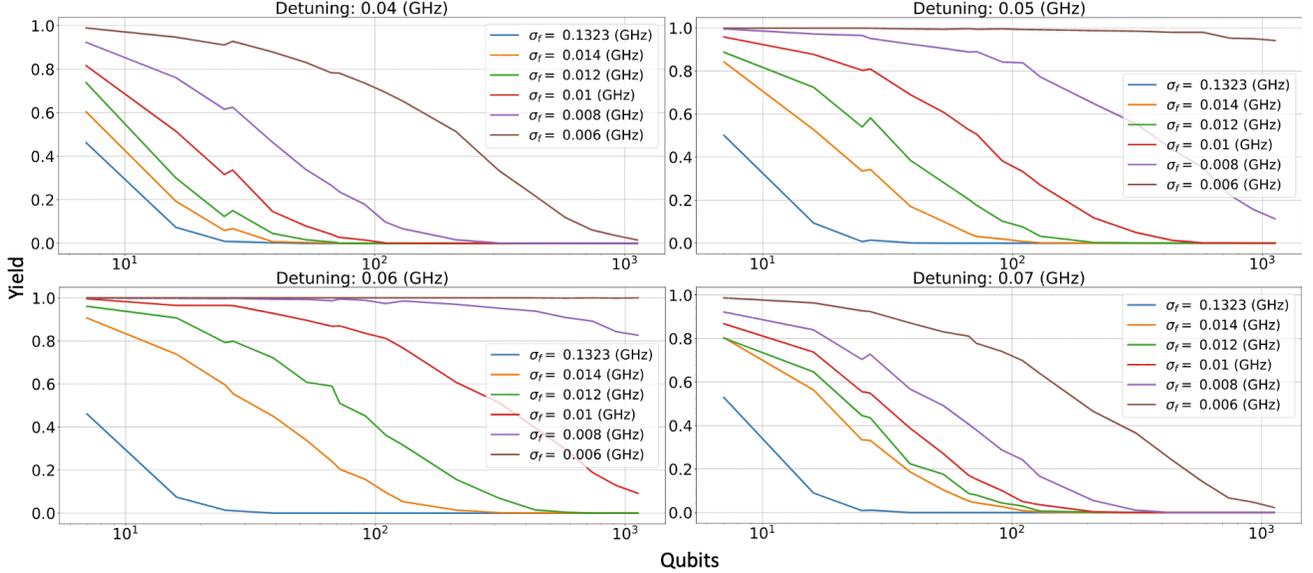}
        \caption{Collision-free yield vs. qubits for detuning values ranging 0.04--0.07 GHz between ideal frequencies, $F_{0,1,2}$.}
        \label{fig:detuning-compare}

\end{figure*}

To gain a better understanding of the relationship between chip size and collision-free yield, we employ a Monte Carlo (MC) model to quantify collision events as heavy-hex devices increase in qubit count. The heavy-hex layout was chosen for analysis because of its use in state-of-art IBM transmon QCs, and current gate error rates are gradually approaching the thresholds required for ECCs on heavy-hex devices. More information can be found in Section~\ref{sect:transmon-qubits}.

Yield simulation involved virtually constructing heavy-hex QCs up to $\sim10^3$ qubits in size. To emulate variation associated with fabrication in our modeling, qubit frequency was assigned by randomly sampling values drawn from normal distributions characterized by both the qubit's ideal frequency, $F_{0,1,2}$ (Section~\ref{sec:arch-constraints}), and fabrication precision, $\sigma_f$ (Section~\ref{sec:device-fab-var}). Next, QCs were evaluated for collisions by checking the criteria in Table~\ref{tab:collision criteria} for satisfiability. If all seven criteria return \texttt{false}, a QC is categorized as collision-free. A batch size of $10^3$ devices was used in MC simulation.

The first step in simulation required determining the optimum frequency detuning for $F_{0,1,2}$ that results in highest yield. Transmon qubits are designed to operate at $\sim 5$ GHz~\cite{malekakhlagh2020first,jurcevic2021demonstration}, therefore $F_0$ was set as 5 GHz.
It should be noted that although detuning between frequencies is important, absolute values are not~\cite{hertzberg2021laser}. In this study we assume step size to be equal between $F_{0,1,2}$ as done in prior work~\cite{li2020towards,hertzberg2021laser}; exploring the impact of varying the distance between ideal frequencies could be an area for future work. 

Fig.~\ref{fig:detuning-compare} shows yield simulation results where each plot employs a different step size ranging from 0.04 to 0.07 GHz between $F_{0,1,2}$. In these plots, the values for $\sigma_f$ were chosen based on the results of~\cite{hertzberg2021laser}, where $\sigma_f=0.1323$ GHz was the qubit frequency spread around targeted frequencies directly after fabrication, $\sigma_f=0.014$ GHz was the fabrication precision obtained via qubit laser tuning, and $\sigma_f=0.006$ GHz, if physically obtainable, is proposed as the threshold for variation to achieve QCs larger than $10^3$ qubits using current collision criteria found in Table~\ref{tab:collision criteria}. The lower left plot of Fig.~\ref{fig:detuning-compare} shows that a target qubit-qubit detuning of 0.06 GHz produces the highest yields during simulation for each level of precision, closely recreating the projections from~\cite{hertzberg2021laser}. This match helps validate our implemented model. Thus, $F_{0,1,2}$ is set as 5.0, 5.06, and 5.12 GHz, respectively, in further analysis as it provides the most optimum frequency step to target during fabrication when considering the Table~\ref{tab:collision criteria} criteria. Fig.~\ref{fig:detuning-compare} demonstrates that yield decreases as frequency step size moves away from 0.06 GHz. The supporting intuition is that if step size is too small, the likelihood increases for post-fabrication frequency detuning to be near-null, producing a Type 1 or 5 collision in Table~\ref{tab:collision criteria}. If step size is too large, however, the likelihood increases of encountering a near half-anharmonicity or near anharmonicity detuning, resulting in Type 3, 4, 6, and 7 collisions. 

The key takeaway from Fig.~\ref{fig:detuning-compare} is that machine variation becomes detrimental as QCs increase in size. When $\sigma_f > 0.006$ GHz and ideal frequency detuning is constant, larger chips experience lower yields due to the increased probability of a frequency collision. 
Today's state-of-art transmon processing has resulted in precision values of 0.0185 GHz~\cite{zhang2020high} and 0.014 GHz~\cite{hertzberg2021laser}. Thus, significant improvements are needed to reach the ideal $\sigma_f<0.006$ GHz that will assist 
QC 
scaling beyond $10^3$ qubits if current collision criteria is considered. Notably, a lower bound for a physically possible $\sigma_f$ will likely be reached as research in the area continues. At this time, the $\sigma_f$ lower bound is unknown and difficult to speculate. 

In the interest of keeping our modeling forward-focused without establishing overly aggressive projections, we set $\sigma_f=0.014$ GHz, the current state-of-art~\cite{hertzberg2021laser}, in our QC system modeling. Some amount of variation in QC fabrication will likely be unavoidable moving into the future, having detrimental impact on the overall fidelity of monolithic QCs. As a result, we are motivated to explore chiplet-based architectures to produce larger machines with greater success.

\begin{figure}[t]
     \centering

         \includegraphics[width=0.97\columnwidth,trim={1.5cm 1.5cm 0cm 0cm},clip]{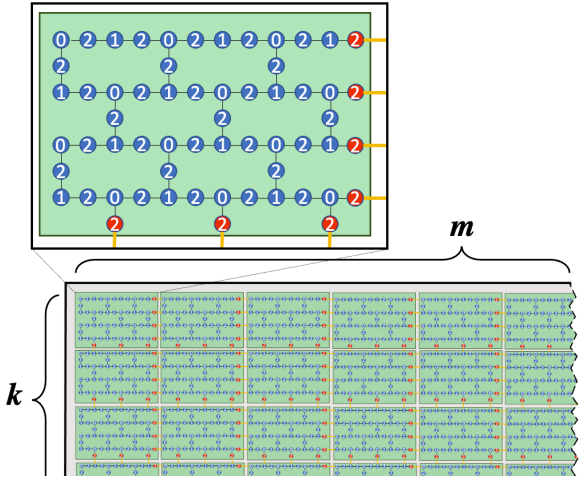}
        \caption{60-qubit chiplet within a $k\times m$ MCM. Links connecting chiplets are indicated in yellow.}
        \label{fig:mcm-example}
\end{figure}

\section{Architecting Quantum Multi-chip Modules}

 Transmon QC yield modeling of Section~\ref{yield-vs-qubits} revealed challenges centered around fabrication variation. The key takeaway is that when monolithic QCs are designed to contain more qubits, the probability of encountering a frequency collision increases. As a solution, we investigate quantum chiplets that achieve higher yields by containing fewer qubits. Chiplets are interconnected with a host chip, creating multi-chip modules (MCMs) as a monolithic alternative.

It should be noted that an inadvertent bonus of MCM architecture, in addition to aformented benefits relating to yield, is that failures within a sub-component have the opportunity to be isolated, preventing widespread errors in computation. For example, the sensitivity of current QCs forces them to suffer from large amounts of correlated error when exposed to random external stimulus such as stray radiation or cosmic rays~\cite{martinis2021saving}. Since each chiplet has some buffer from the active components of its neighbors, large-scale qubit corruption from electromagnetic contamination can be avoided. 

\subsection{Chiplet-based Design} 
\label{sec:chiplet-based-design}

\textbf{Chiplet Design: }A chiplet containing 20 qubits was first introduced in Table~\ref{tab:collision criteria}, column 5. This device was designed to include a single, complete heavy-hex honeycomb -- other chiplet designs include a partial heavy-hex ring or multiple. Each chiplet has edge qubits, indicated in gray, that support coupling to qubits on adjacent chips. The heavy-hex three-frequency pattern is described in the Table~\ref{tab:collision criteria}, column 5 drawing. 
 When linked to other devices using inter-chip links, the design of the chiplet allows the heavy-hex lattice and frequency allocation pattern to be preserved in the resulting MCM, keeping future error correction with the hybrid surface/Bacon-Shor code in mind. 

An example of a 60-qubit chiplet is pictured within a $k\times m$ MCM in Fig.~\ref{fig:mcm-example}.  This device's structure closely matches that of the 20-qubit chiplet in Table~\ref{tab:collision criteria} but with two additional dense rows (i.e. rows containing qubits labeled as 0, 1, and 2) holding four extra qubits each and two additional sparse rows (i.e. rows that only contain qubits labeled as 2) holding one additional qubit each. Larger heavy-hex chiplets require additional rows and columns that are added similarly. 

\textbf{MCM Interconnects: } 
Recently, SC qubits on separate chips were coupled with high fidelity using microwave link technology~\cite{zhong2021deterministic,magnard2020microwave}. In another demonstration highly relevant to this work, SC qubits on two separate physical modules were flip-chip bonded to a larger carrier chip, or interposer~\cite{gold2021entanglement}. Inter-chip coupling rates and entanglement quality comparable to intra-chip qubit-qubit coupling were accomplished. Although either linking technique could be applied toward networking an MCM, here, we focus on the latter work that makes progress in flip-chip architectures. 

In~\cite{gold2021entanglement}, the average coherence limited fidelity, or the maximum achievable fidelity predicted from the measured relaxation and dephasing times, of the two-qubit operations across module links was 92.5\% while the median was 94.4\%. These statistics show that the cross-module gates are more error-prone than the on-chip operations of the example IBM machines in Fig.~\ref{fig:generation-comparison}(a). The difference in fidelity, however is not drastic, demonstrating that flip-chip bonding holds more than just theoretical viability when applied to QCs.

Inter-chip links of the Fig.~\ref{fig:mcm-example} MCM are indicated in yellow. Since the right-most and bottom-most qubits in our chiplet design always have a $F_2$ assignment, they will act as the control in inter-chiplet $CR$ interactions.  

\begin{figure}[t]
     \centering
         \includegraphics[width=0.97\columnwidth,trim={2.1cm 0cm 2cm 0cm},clip]{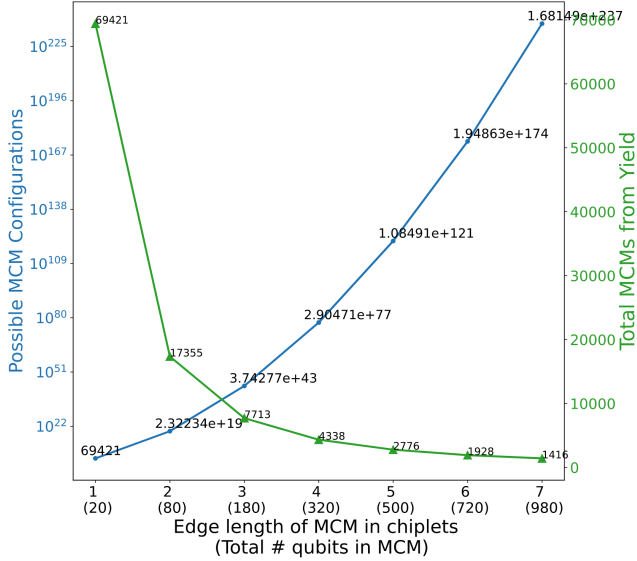}
        \caption{Potential configurations for a single MCM (left axis) and upper bound of number of assembled MCMs (right axis) vs. MCM size. Square MCMs ($m\times m$) use 20-qubit chiplets. Calculations based on $\sim 69.4\%$ yield with $\sigma_f = 0.014 $ GHz and a batch size of $10^5$ units.}
        \label{fig:mcm-details-20q-chip}
\end{figure}

\subsection{Customization and Reconfigurability}
\label{sec:MCM-customization-reconfig}

One of the most notable advantages of a chiplet architecture is flexibility. For instance, the dimensions of the MCM can be chosen ad hoc. When the MCM increases in total number of chiplets selected from the collision-free yield, the amount of possible system configurations grows at a factorial rate. This relationship of configurations vs. MCM size is seen on the left axis in Fig.~\ref{fig:mcm-details-20q-chip}. This plot shows configuration totals based on $\sim 69.4\%$ yield of identically-designed, 20-qubit chiplets, resulting from a state-of-art $\sigma_f = 0.014 $ GHz and a QC batch size of $10^5$ units. The influence of a fabrication precision of $\sigma_f = 0.014 $ GHz on the collision-free yield of QCs up to $10^3$ qubits is shown by the orange curves in Fig.~\ref{fig:detuning-compare}.

The right half of plot in Fig.~\ref{fig:mcm-details-20q-chip} describes the total number of MCMs that result from the collision-free yield of 69,421/100,000 chiplets. As expected, the total number of MCMs resulting from the collision-free subset decreases as dimension increases, but there still exists the opportunity for a large amount of customization. We assume use of the industry-standard known good-die (KGD) testing techniques~\cite{kannan2015enabling} where individual chips are tested before MCM assembly. Thus, QC chiplets are sorted in a process similar to speed-binning: devices with similar fidelity properties can be post-selected and linked, resulting in systems with uniform and high-fidelity quantum computation. 

Fig.~\ref{fig:mcm-details-20q-chip} considers MCMs composed of chiplets from the same batch. However, innovation in quantum processing, control, and carrier-chip methods could result in further MCM customization (e.g., mixed integration of heterogeneous physical qubit technologies). Highly specialized systems comprised of multiple types of physical qubits could present new opportunities for application-specific quantum accelerators. 

\subsection{Monolithic vs. MCM Fabrication Output}
\label{sec:mono-vs-mcm-output}

Section~\ref{yield-vs-qubits} demonstrates the collision-free yield benefits of smaller transmon QCs. Here, we explore how yield improvements can translate into significantly more fabrication output, and thus greater cumulative compute power, via MCM systems. We build an analytical model that supports the simple argument for chiplets presented in Fig.~\ref{fig:chiplet-vs-monolithic}. The guiding principle is that chiplets exploit the ability to process more devices at once since their die takes less area on a wafer.

To build our model, the collision-free yield associated with a $q_m$-qubit monolithic QC is defined as $Y_m$ out of a batch size of $B$ die. Conversely, $Y_c$ is the yield associated with smaller, $q_c$-qubit chiplets. If the chiplets are designed for integration within an $k\times m$ MCM, the total number of resulting MCMs, $N$, is calculated as 

\begin{equation}
    N = \frac{Y_c\times ( B \times \frac{q_m}{q_c})}{k\times m} 
\label{eq:upper-bound}
\end{equation}

\noindent where the die area ratio between the monolithic and chiplet QCs is approximated using their respective qubit capacity, $\frac{q_m}{q_c}$. 

The yield simulation of Fig.~\ref{fig:detuning-compare} is applied to show how the upper bound of assembled MCMs compares to the yield of similarly-sized monolithic systems. Consider the experiment where $\sigma_f = 0.014$ GHz and the frequency detuning is $0.06$ GHz in the lower left plot of Fig.~\ref{fig:detuning-compare}. With these parameters, a $q_m = 100$ QC has a yield of approximately $Y_m = 0.11$ for $B = 1000$ chips. On the other hand, a $q_c=10$ chiplet is characterized by approximately $Y_c = 0.85$. Using the same wafer area as the monolithic device, $B \times \frac{q_m}{q_c} = 1000\times \frac{100}{10} = 10,000$ chiplets theoretically could be fabricated, where we would expect 8,500 as collision-free. MCMs of dimension 2$\times$5 enable ten-chiplet QCs that contain the same number of qubits as the monolithic devices, thus an $N = 850$ upper bound of 100-qubit MCMs would result. As compared to the $Y_m \times B = 0.11 \times 1000 = 110$ device yield from 100-qubit monolithic production, MCMs result in $\sim7.7\times$ gain in manufactured QCs.  

\subsection{Practical Constraints of MCMs}
\label{practical-constraints-MCMs}

Section~\ref{sec:mono-vs-mcm-output} describes the upper bound for MCM fabrication output as compared to monolithic devices. This provides an estimate for potential gains in fabrication output when taking a modular QC approach, but reaching this upper bound is unrealistic. In reality, some MCMs will be lost during carrier-chip bonding procedures due to link faults or other defects that prevent practical use. Thus, it is important that assembly yield is considered. Our proposed architecture unites quantum chiplets through flip-chip bonding to a larger, carrier chip, or interposer. Here, we will include MCM assembly yield to account for inevitable failures during assembly. 
Our model is based on silicon interposer defect rates in~\cite{kannan2015enabling}: we estimate the success rate for the placement of a controlled collapse chip connection (C4) bump bond on a passive carrier chip. Fabrication information from~\cite{gold2021entanglement} provides insight about the number of bump bonds required for each inter-chip linked qubit.   
Future work will refine the MCM assembly yield as data for fabricated quantum devices becomes available.

\section{Framework for Quantum MCM Modeling}

Potential benefits of scaling via quantum MCMs include increased yield, more homogeneous computation, and added flexibility. The practicality of these systems, however, must be evaluated. 
In this section, we present our MCM modeling framework. 
A corresponding tool integrates these methods for simulating QC yield and estimating device and workload performance. 
As a note, while parameters in our analysis are based on state-of-art transmon~\cite{IBMQS} and flip-chip technology~\cite{gold2021entanglement}, properties involving device coupling, frequency assignment, fabrication precision, and inter- and intra-chip gate error are parameterized to accommodate future QC improvements.

\subsection{On-chip Fidelity Assignment}
\label{sect:Fidelity-on-chip}

\begin{figure}[t]
     \centering
         \includegraphics[width=0.97\columnwidth,trim={0cm 2.8cm 0cm 3cm},clip]{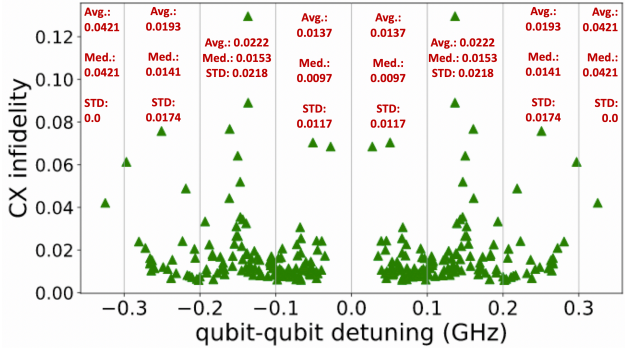}
        \caption{Plot of $CX$ infidelity vs. qubit-qubit detuning for 127-qubit IBM Washington, an Eagle-class processor. Each datapoint represents an average of characterization data for a qubit-qubit connection over 15 calibration cycles. Median of plotted points is 0.012; average is 0.018.}
        \label{fig:cx-infid-detuning}
\end{figure}

Section~\ref{yield-vs-qubits} describes collision-free yield. By avoiding Table~\ref{tab:collision criteria} conditions, QCs theoretically have $\lesssim 1\%$ frequency-related error contributing to gate infidelity. Real quantum systems, however, are characterized by diverse noise profiles that also contribute to gate error. Examples of two-qubit gate infidelity statistics for three QCs are found in Fig.~\ref{fig:generation-comparison}(b). To better capture the complexities of real quantum operations in our modeling, we employ real QC data in our analysis.

The $CR$ gate, Section~\ref{sect:CR-gate}, along with single-qubit rotations generate the $CX$ operation~\cite{corcoles2013process}. 
$CX$ fidelilities 
and qubit frequencies are available in the IBM QC backend properties. Our model uses this information to develop an emperical relationship between qubit-qubit detuning and two-qubit operator fidelity. In our study, the QC under investigation is IBM Washington, a state-of-art Eagle-class processor that accomplished an quantum-industry first of surpassing the 100-qubit milestone~\cite{eagle-processor}. IBM Washington's innovation in transmon control and design represents the future goals of quantum technology, so its properties were used to assign two-qubit gate fidelity values on-chip. 

Real QC noise fluctuates over time~\cite{dasgupta2021stability},
thus we gathered and averaged $CX$ infidelity over 15 calibration cycles 
and correlated the averages with the frequency detuning of the involved qubits. Fig.~\ref{fig:cx-infid-detuning} plots the relationship of average $CX$ infidelity vs. detuning for IBM Washington. The median of plotted data is $0.012\rightarrow1.2\%$ while average is $0.018\rightarrow1.8\%$. As illustrated in Fig.~\ref{fig:cx-infid-detuning}, data was binned according to detuning intervals of step-size 0.1 GHz. This interval was inspired by the observed fabrication-induced frequency spread described in~\cite{zhang2020high}, but the parameterized nature of the presented modeling framework allows the sampling bounds to be adjusted. After qubit-qubit detuning characterization, gate fidelity is assigned by sampling from the distribution of the corresponding bin. 

\subsection{Inter-chip Link Fidelity Assignment}
\label{sec:link-assignment}

An important consideration for MCM architectures is the trade-off between chiplet yield and inter-chip communication. In our MCM architecture, chiplets are proposed to be flip-chip, C4 bump bonded to an interposer. Successful two-qubit operations spanning QC modules have been experimentally demonstrated with fidelity averaging 92.5\% (Section~\ref{sec:chiplet-based-design}). Our model assigns link infidelity based on the distribution of~\cite{gold2021entanglement}. As a note, the link vs. on-chip average error ratio is $e_{link}/e_{chip}=$ 7.5\% / 1.8\% $\approx$ 4.17.

\section{Architecture Evaluation}

This section examines MCM and monolithic trade-offs in terms yield, average two-qubit gate infidelity, and application performance. Evaluation will be completed using devices within collision-free yields that are modeled using properties of state-of-art transmon QCs and quantum link hardware. In the absence of quantum chip interposer data, MCM assembly yield will be based upon silicon interposer defect rates~\cite{kannan2015enabling} due to the parallels between classical and emerging quantum flip-chip bonding techniques~\cite{gold2021entanglement}.

\subsection{Benchmarks}
\label{sec:benchmarks}

Seven benchmarks were used to evaluate the MCM and monolithic systems. The circuits were chosen to adequately cover the application space of realistic QC workloads~\cite{tomesh2022supermarq,teague-benchmarks}. Circuits were designed
for 80\% system qubit utilization to allocate ancilla for compiler mapping and optimization.  

\textbf{Bernstein-Vazirani (BV):}
 demonstrates a quantum advantage over classical programming by guaranteeing the return of the bitwise product of some input with a hidden string~\cite{bernstein1997quantum}. 

\textbf{Quantum Approximate Optimization Algorithm (QAOA):} 
a hybrid quantum-classical algorithm for applications solving NP-Hard combinatorial optimization problems~\cite{farhi2014quantum}. 

\textbf{Greenberger–Horne–Zeilinger (GHZ):} 
prepares large-scale entanglement that is required by many complex quantum algorithms and communication protocols for quantum-enabled computational speedups~\cite{greenberger1989going}.  

\textbf{Ripple-Carry Adder:} a critical subroutine in quantum algorithms such as Shor's quantum factoring. We include the implementation described in~\cite{cuccaro2004new}.

\textbf{Quantum Primacy:} generates random quantum circuits similar to those proposed for and used to demonstrate quantum primacy, or advantage~\cite{boixo2018characterizing,arute2019quantum}.

\textbf{Bit Code:} implements a syndrome measurement in a bit-flip ECC.

\textbf{Hamiltonian:} constructs circuits that simulate 1D Transverse Field Ising Models (TFIM) used to discover static properties of quantum systems, such as ground-state energies of molecules~\cite{bassman2020towards}.

\subsection{Methodology}
\label{sec:methodology}

Variation during fabrication, on-chip gate error, and link quality were set using state-of-art QC data. We anticipate future improvements will scale down infidelity as qubits scale up. 
Here, our focus is on the near term, and our analysis includes QCs comprising of several hundred qubits.

We considered chiplets with 10, 20, 40, 60, 90, 120, 160, 200, and 250 qubits. We evaluated a total of 102 MCMs against their monolithic counterparts. MCM dimensions of $k\times m$ were chosen so that each MCM in a chiplet category had a unique size $\leq 500$ qubits. Since duplicates were avoided in terms of number of qubits, MCM dimensions that were more `square' were prioritized to reduce topology graph diameter. For example, a 40-qubit MCM of dimension $2\times 2$ with 10-qubit chiplets was included in our analysis whereas a $4\times 1$ configuration with 10-qubit chiplets was omitted.

Yield information resulted from MC simulation with $F_{0,1,2}$ set to 5.0, 5.06, and 5.12 GHz (Section~\ref{yield-vs-qubits}), $\sigma_f = 0.014$ GHz (fabrication precision of~\cite{hertzberg2021laser}), and an original batch size of 10,000 units. After Table~\ref{tab:collision criteria} criteria evaluation, collision-free chiplets were grouped for MCM assembly.

MCM assembly creates as many high-quality MCMs as possible. The exploration space for the best chiplets to use within an MCM is exponential, and rather than search for optimums or use random selection from the collision-free bin, infidelity and frequency information gathered from prior KGD characterization guides MCM assembly. Chiplet stitching procedures use the chiplets with the lowest error rates first and the highest error rates last from the collision-free bin. As many complete and collision-free MCMs possible are created following the principle that once a chiplet is within an MCM, it is removed from the collision-free bin. If a frequency collision between adjacent chiplets is found with a particular MCM configuration, chiplet placement is shuffled within the MCM. If a collision-free MCM is not discovered according to time-out criteria selected during runtime (100 maximum reconfigurations), chiplets are returned back to the bin and MCM assembly continues with a new subset of chiplets from the sorted, collision-free bin. 

Yield analysis includes loss due to the inability to place chiplets within collision-free MCMs and linking faults. Linking faults during assembly describe the failure to properly bond linking qubits on the chiplet edge to the carrier chip so that they can be coupled inter-chip. We used data involving silicon interposer defect rates from~\cite{kannan2015enabling} to derive a success probability estimation for each C4 bump bond on a passive carrier chip of $s_l=$99.999960642\%. Fabrication information from~\cite{gold2021entanglement} inspired the allocation 25 bump bonds for each linked qubit on a chiplet's edge. Thus, each qubit linked inter-chip has a $s_l^{25}$ probability of a successful bond.

During infidelity analysis, two-qubit gate infidelity for collision-free devices was assigned using real-machine data 
(Section~\ref{sect:Fidelity-on-chip}). 
Average infidelity averaged across every qubit pair is assumed as known and used to rank chiplets  
by performance potential to 
 prioritize the use of the best chiplets within MCM construction (Section~\ref{practical-constraints-MCMs}). 
Once assembled and flip-chip bonded, we assigned each MCM's two-qubit gate error over inter-module links (Section~\ref{sec:link-assignment}). 
We studied the influence of MCM architectures on overall system average two-qubit operation infidelity.

For application analysis, the benchmarks described in Section~\ref{sec:benchmarks} were generated to target 80\% of qubits on the MCM and monolithic devices. 
Application evaluation used fidelity product of gates, a figure of merit based on the estimated probability of success (ESP) metric frequently used in prior work to predict quantum program success on hardware~\cite{tannu-diverse-mappings,murali2020architecting,nishio2020extracting,li2022paulihedral}. Since our simulation methods include the assignment of two-qubit infidelity between intra- and inter-chip connected qubits post fabrication and assembly, respectively, our fidelity product that estimates benchmark success is calculated by multiplying all two-qubit operator fidelities. 
We note that some higher-dimension MCMs lack a monolithic counterpart during evaluation due to 0\% yield during MC simulation. 

\begin{figure}[t]
     \centering
         \includegraphics[width=0.97\columnwidth,trim={0.05cm 0.8cm 0.2cm 1cm},clip]{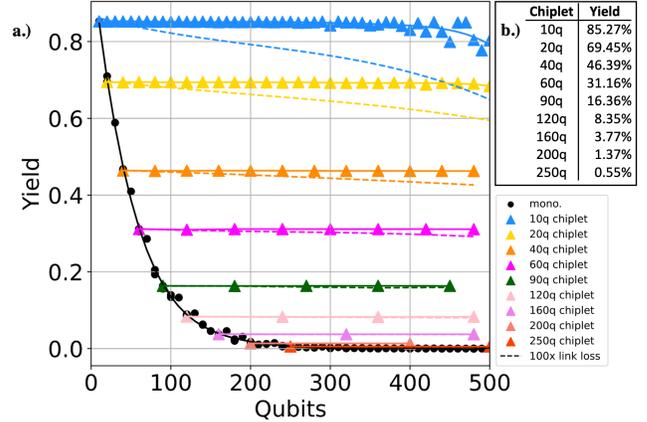}
        \caption{(a) Plot of yield vs. qubits. Curves represent different architecture types: monolithic and MCM. MCM yield includes yield losses from assembly and linking.  
        Yield vs. qubits relationship where link qubit bonding failure is 100$\times$ is indicated with dashed line. (b) Chiplet yields. 
        }
        \label{fig:qubits-yield}
\end{figure}

\subsection{Results}

In this section, yield, gate performance, and application-based
analysis show the feasibility of QC scaling via modular design. Data used to evaluate quantum MCM and monolithic architectures can be found in~\cite{chiplet-repository}.

\begin{figure*}[!th]
     \centering
         \includegraphics[width=0.97\textwidth,trim={0cm 7cm 0cm 7cm},clip]{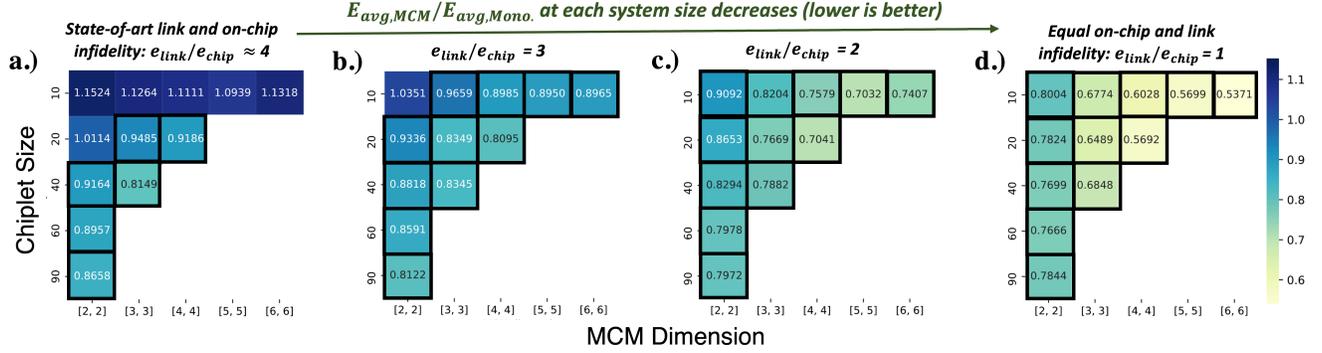}
        \caption{Heatmaps showing the MCM vs. monolithic ratio of infidelity averaged across every qubit pair, $E_{avg.,MCM}/E_{avg.,Mono}$ using the scaled collision-free yields. A ratio less than one indicates that the MCM systems have lower average two-qubit gate infidelity. Modeling includes (a) state-of-art link and on-chip infidelity, $e_{link}/e_{chip}\sim 4$, along with cases where links in MCMs have improved error: (b) $e_{link}/e_{chip} = 3$, (c) $e_{link}/e_{chip}= 2$, and (d) $e_{link}/e_{chip} = 1$ (on-chip and link two-qubit gate infidelity equal).}
        \label{fig:heatmaps}

\end{figure*}

\subsubsection{Yield Evaluation}

We first evaluate MCM vs. monolithic QC architectures in terms of yield. 
Yield is an important consideration in QC design: although it might be possible to fabricate a high-quality, monolithic QC of a certain qubit size, the trade-off of scrapped chips with critical defects (i.e. frequency collisions) produced along with one good chip might be prohibitively expensive. Yield analyisis results are found in Fig.~\ref{fig:qubits-yield}. A plot of yield vs. system qubits for monolithic QCs and MCM QCs constructed with varied chiplet size is featured in Fig.~\ref{fig:qubits-yield}(a). It should be noted that the yield data for each system type is not strictly monotonic due to the stochastic nature of MC simulation and the slight influence that QC dimension has on the occurrence of frequency collisions (i.e. more ``square'' devices have more qubit-qubit communication) and thus yield. The MCM curves consider loss during assembly caused by 1) chiplets that cannot be placed in a complete, collision-free MCM and 2) link failures. Post-assembly yield is calculated as the the ratio of total chiplets used to assemble complete, collision-free MCMs vs. total chiplets in the original fabrication batch (i.e. 10,000) multiplied by $(s_l^{25})^L$, where L is the total number of edge qubits used for inter-chip MCM linking. To show sensitivity to link qubit loss during assembly, dashed lines in~\ref{fig:qubits-yield} show adjusted yield when the probability of bonding failure is 100$\times$. Regardless of the amplified failure during assembly, MCM systems show considerable benefits in terms of yield. Monolithic systems, however, show the most dramatic drop in yield, dropping to $\sim10\%$ at 120 qubits, alluding that monolithic architectures are the most costly in terms of wasted resources during fabrication. Chiplet yields are explicitly stated in Fig.~\ref{fig:qubits-yield}(b). These serve as the upper bound of MCM yield if all elements of the collision-free chiplet bins can be successfully placed within a collision-free MCM. The assembly and linking contributions to yield cause the plotted points of the MCM curves to fall below the yields of Fig.~\ref{fig:qubits-yield}(b). We note, however, that our results find that that the assembly and linking loss only slightly impact yield benefits of the MCM sizes under investigation. 

A key takeaway from Fig.~\ref{fig:qubits-yield} is that with $\sigma_f = 0.014$ GHz, it is unfeasible to create a monolithic device $\gtrsim400$ qubits because of prohibitively low yield. As devices approach 500 qubits, MCM architectures have slightly lower yields than their base-chiplet yields described in Fig.~\ref{fig:qubits-yield}. This is due to the increased likelihood of encountering a defect during assembly as more inter-chiplet links are required. Additionally, a greater number of leftover chiplets result that cannot be placed in a complete, collision-free MCM of larger dimension. Regardless, MCM yields are still typically higher than that of the monolithic QCs. 
All configurations considered, MCM groups based on the 10-, 20-, 40-, 60-, 90-, 120-, 160-, and 250-qubit chiplets demonstrate $9.58\times$, $11.34\times$, $13.45\times$, $16.53\times$, $23.03\times$,
$30.44\times$,
$92.61\times$, and
$52.99\times$
average yield improvement relative to their monolithic counterparts, respectively. The 200-qubit chiplet was excluded from this analysis since it only creates a 400-qubit MCM that had a corresponding monolithic device with 0\% yield during MC simulation. Theoretically, MCM yield improvement is infinite when monolithic yields are 0\%.

\subsubsection{Infidelity Evaluation}
\label{sec:infidelity-eval}

We compare the influence of chiplet size on average infidelity averaged across every qubit pair, $E_{avg.}$, for the MCM and monolithic systems. We also evaluate the relationship between $E_{avg.,MCM}$ and $E_{avg.,Mono}$ as average link error, $e_{link}$, improves.

To focus this analysis, we evaluate a subset of our experimental systems where the MCMs are $n\times n$ (i.e. chiplets are assembled in a `square'). Square structures are favorable because they reduce the graph diameter of the system topology, minimizing worst-case communication distance between qubits. Fig.~\ref{fig:heatmaps} shows heatmaps describing the MCM vs. monolithic average infidelity averaged across every qubit pair, $E_{avg.,MCM}/ E_{avg.,Mono}$ for various link quality conditions. A ratio less than one (magnitude indicated by cell brightness) indicates that the MCM has lower average error rates as compared to its monolithic counterpart. These areas within the heatmap are highlighted with a thick, black border. It is important to reemphasize that yield benefits of smaller chiplets often enable the production many more MCMs than monolithic devices of equivalent size. Thus, the yield relationship scales the number of MCMs used in architecture comparison: we compare the devices in the collision-free monolithic yield to the MCMs resulting from the chiplets in the scaled, collision-free bin.

The $E_{avg.,MCM}/E_{avg.,Mono}$ ratios of Fig.~\ref{fig:heatmaps}(a) analyze systems that use state-of-art inter-chip link and on-chip infidelity. As described in Section~\ref{sec:link-assignment}, $e_{link}/e_{chip} =$ 7.5\% / 1.8\% $\approx$ 4, prohibiting MCMs with 10-qubit chiplets from showing an advantage in terms of average infidelity, even if they have considerable yield benefits. Small chiplets are typically characterized by lower two-qubit error rates, especially those that rank higher in the sorted, collision-free bin. However, the number of links that the smaller-chiplet MCMs require inflate average error rates. 
With chiplets of 20 qubits and larger, we begin to see MCMs surpass the performance of the monolithic systems. 
For example, the 3$\times$3 MCM with a 20 qubit chiplet (180 qubits) has $E_{avg.,MCM}/E_{avg.,Mono}=$ 0.9485 and the 2$\times$2 MCM with a 40 qubit chiplet (160 qubits) has $E_{avg.,MCM}/E_{avg.,Mono}=$ 0.9164. Our MCM modeling that uses state-of-art error values show improvements where MCM average two-qubit gate infidelity is reduced to be about 81.5\% of its monolithic counterpart for the case of the 40-qubit chiplet arranged in a $3\times 3$ MCM.

In the future, it is projected that the quality of communication between separate chiplets coupled on an carrier interposer a will improve as engineering challenges are overcome~\cite{gold2021entanglement}. In the extreme case where average link error rates are equivalent to on-chip error rates, pictured in Fig.\ref{fig:heatmaps}(d), MCMs outperform monolithic devices in terms of $E_{avg}$ in 100\% of considered system configurations. However, achieving equivalent on-chip and link qubit-qubit coupling is unnecessary for MCM overall system infidelity, in terms of $E_{avg}$, to be less than its monolithic counterpart. We find at $e_{link}=2e_{chip}$, all $E_{avg.,MCM}/E_{avg.,Mono}$ ratios are $<1$.

\begin{figure*}[t]
     \centering
         \includegraphics[width=0.97\textwidth,trim={.1cm 3.7cm 0.1cm 3.5cm},clip]{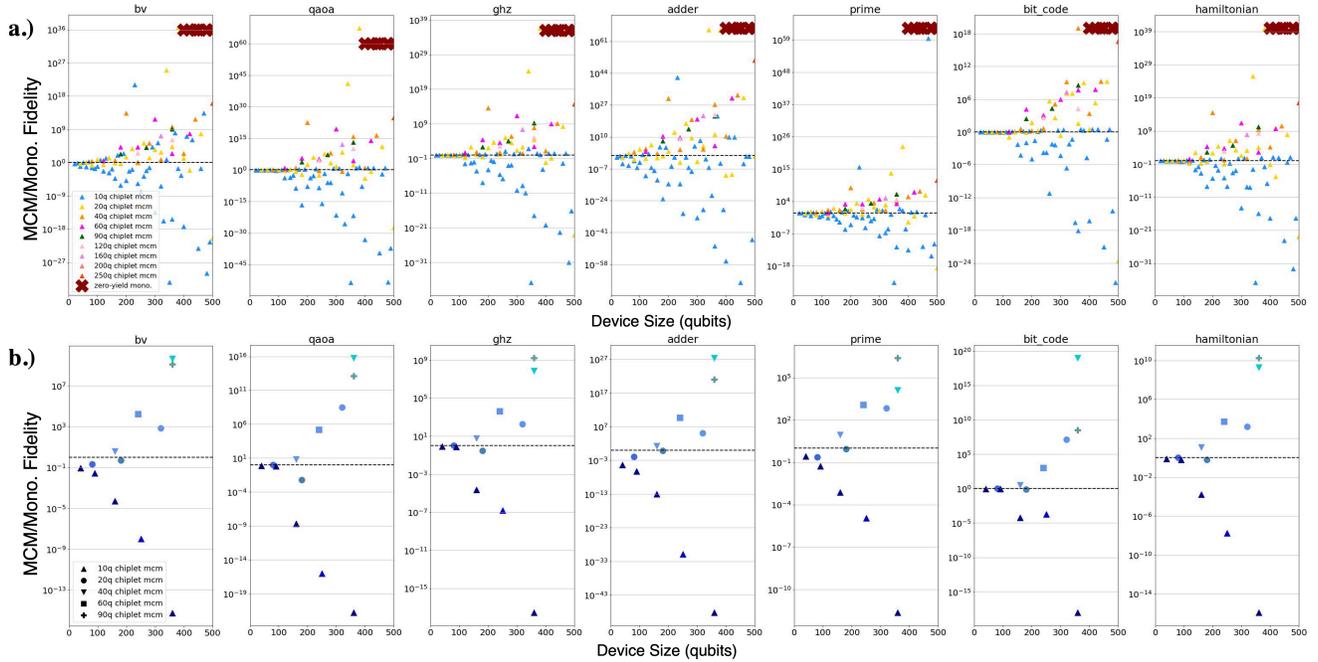}
        \caption{MCM and Monolithic QC benchmark evaluation. (a) Results for all evaluated systems and (b) Results for `square' systems featured in Fig.~\ref{fig:heatmaps}(a)}
        \label{fig:benchmark-results}

\end{figure*}

\begin{table}
\begin{center}

\begin{tabular}{ c | c | c | l }
\textbf{Chip.} & \textbf{Dim.} & \textbf{Qubits} & \textbf{Bench:1q / 2q / 2q critical}\\
\hline
10q & 2x2 & 40 &bv: 192 / 437 / 228\\ 
& & & g: 3 / 31 / 31 \\
& & &q: 34 / 62 / 33 \\
& & & a: 390 / 1339 / 750 \\
& & &p: 226 / 315 / 74 \\
& & &  bc: 16 / 30 / 30\\
& & & h: 191 / 62 / 62\\
\hline
20q & 2x2 & 80 & bv: 384 / 1193 / 604\\
& & &  g: 3 / 63 / 63\\
& & & q: 66 / 126 / 65\\
& & &  a: 806 / 3741 / 1881\\
& & & p: 461 / 890 / 88\\
& & &  bc: 32 / 62 / 62\\
& & & h: 383 / 126 / 126\\
\hline
40q & 2x2 & 160 & bv: 768 / 2251 / 1135\\
& & & g: 3 / 2170 / 1147\\
& & & q: 130 / 3691 / 1263\\
& & & a: 1638 / 6915 / 3695\\
& & & p: 927 / 1552 / 114\\
& & & bc: 64 / 2557 / 1313\\
& & & h: 767 / 2320 / 1267\\
\hline
60q & 2x2 & 240 & bv:   1152 / 5865 / 2952\\
& & & g: 3 / 6081 / 3105\\
& & & q: 195 / 9320 / 3245\\
& & & a: 2470 / 16164 / 7654\\
& & & p: 1380 / 4690 / 300\\
& & & bc: 96 / 3804 / 1973\\
& & & h: 1151 / 6241 / 3235\\
\hline
90q & 2x2 & 360 & bv: 1728 / 6393 / 3282\\
& & & g: 3 / 5977 / 3112\\
& & & q: 291 / 10410 / 3413\\
& & & a: 3717 / 19682 / 9818\\
& & & p: 2073 / 4382 / 205\\
& & & bc: 144 / 6016 / 3133\\
& & & h: 1727 / 6455 / 3480
\end{tabular}%

\end{center}
\caption{Details of select compiled benchmarks.}
\label{tab:benchmark-info}
\end{table}

\subsubsection{Application Evaluation}

Fig.~\ref{fig:benchmark-results} shows MCM and monolithic QC performance when BV, QAOA, GHZ, adder, quantum primacy, bit-code, and Hamiltonian simulation workloads (Section~\ref{sec:benchmarks}) were mapped to the systems under investigation. Since our study analyzed hundreds of compiled benchmarks, details for select benchmarks, those used in the Fig.~\ref{fig:heatmaps}(a), column 1 systems, were included in Table~\ref{tab:benchmark-info} to provide insight about the gate composition and critical paths of example circuits.   
The structures we evaluate surpass the capacity of today's most powerful quantum simulators, so our figure of merit is the fidelity product of all two-qubit gates (Section~\ref{sec:methodology}). Since we are interested in comparing fidelity of the two architectures, the y-axis of the 14 plots in Fig.~\ref{fig:benchmark-results} measures the ratio between MCM and monolithic fidelity. Fig.~\ref{fig:benchmark-results}(a) includes all considered MCM/monolithic devices while Fig.~\ref{fig:benchmark-results}(b) only includes the `square' systems featured in the average infidelity analysis described in Section~\ref{sec:infidelity-eval}, Fig.~\ref{fig:heatmaps}(a).   A value above one indicates a MCM advantage whereas a value below one indicates a monolithic advantage. The distance away from one provides insight about the extent that one architecture outperforms the other. A red ``X'' (11 total on each Fig.~\ref{fig:benchmark-results}(a) plot) indicates where monolithic devices encounter a 0\% collision-free yield. Here, the MCM device offers the only solution for implementing the selected benchmark, theoretically resulting in an infinite MCM/monolithic fidelity ratio. It should be highlighted that the extreme values seen on the y-axis are caused by on-chip and inter-chip infidelity of magnitude $10^{-2}$. 

A significant takeaway from Fig.~\ref{fig:benchmark-results} is that MCM advantage appears in select cases -- chiplet size and MCM dimension likely influence function. 
 For example, as devices increase in size, the amount of variation also increases, negatively impacting overall device performance. Postselected chiplets can create higher-fidelity clusters within MCMs, but there are trade-offs between chiplet size and link infidelity due to communication needed between MCM modules. It is possible, however, to select MCM configurations such that higher link errors are balanced and their impact on benchmark performance is reduced. 
 While Fig.~\ref{fig:benchmark-results} shows the viability of MCMs, the MCM is not always the highest-performing architecture, especially in smaller systems. This outcome is not unique to our study; past work in classical chiplets have also reported performance losses in modular systems, but often these losses are heavily outweighed by the benefits that manufacturing chiplets over monolithic devices brings~\cite{kim2007architectural}. However, we would like to emphasize that our results show the potential for guiding future quantum architecture design, especially with the results of Fig.~\ref{fig:benchmark-results}(b). Here, the points are colored according to their assignment within the heatmap in Fig.~\ref{fig:heatmaps}(a), and it is shown that MCMs with lower $E_{avg.,MCM}/E_{avg.,Mono}$ tend to have better benchmark performance as compared to their monolithic counterparts. We observe a relationship between the MCM vs. monolithic infidelity (lower better) and the MCM vs. monolithic application fidelity (higher better). In Fig.~\ref{fig:benchmark-results}(b) all of the 40-, 60-, and 90-qubit chiplets show advantage, and their values in Fig.~\ref{fig:heatmaps}(a) are $<1$.

\section{Future Scaling Considerations}

As modular quantum architectures mature, chiplet and MCM specific compilers, perhaps ones based on circuit cutting approaches~\cite{tang2021cutqc}, will be essential for informed use of modular systems through awareness of error rate distributions. Intelligent compilation routines that consider links would allow improved global mapping and optimization.

The proposed chiplets maintain heavy-hex connectivity, supporting eventual adoption of the hybrid surface/Bacon-Shor ECC. However, errors associated with the flip-chip linking technology should be characterized to see if alternative ECCs should be targeted. Dynamic ECC compilation could be explored to allow for adaptive code distances across lower fidelity or more varied sections of the MCM network. 

Last, we hope to improve our modeling by applying first principles~\cite{malekakhlagh2020first} to better capture the relationship between frequency detuning and operator fidelity during analysis. 
Additionally, we hope to refine the modeling of quantum MCM assembly as data becomes available.

\section{Related Work}

Advances in quantum hardware~\cite{eagle-127q,IBM-hardware,chamberland2020topological}, algorithm compilation~\cite{murali2019noise,li2019tackling,shi2019optimized,murali2020software}, and control mechanisms~\cite{baum2021experimental,carvalho2021error} boost the performance of near-term QCs, improving their viability for scaling. Methods to better characterize and reduce quantum defects that impact coherence~\cite{nersisyan2019manufacturing,place2021new,wang2022hexagonal} are also relevant as they aim to make qubits better-suited for future, large-scale computation. Studies are emerging that model linked quantum devices to better understand distributed QC performance in SC-based systems~\cite{laracuente2022short} and trapped-ion devices~\cite{monroe2014large}. Finally, work exists that outlines architecture considerations for teleportation-based quantum multi-core systems~\cite{rodrigo2020will} and explores latency and scaling trade-offs needed in a technology-agnostic manner for effective communication and computation in such systems~\cite{rodrigo2021double,rodrigo2021modelling}.  In this section, we describe how our contributions differ from prior art related to this study. 

The frequency allocation problem associated with transmon QCs along with the yield impact of frequency collisions appeared in prior work~\cite{brink2018device,morvan2022optimizing}. The work of~\cite{hertzberg2021laser} provides a solution to minimize collision conditions by presenting a technique that adjusts transmon frequencies post-fabrication with laser annealing to significantly boost transmon QC yield. Afterwards, laser annealing was successfully applied on real IBM QCs~\cite{zhang2020high} currently deployed on the IBM Quantum Cloud~\cite{IBMQE}. This prior work, along with the fabrication data of the experimental paper~\cite{gold2021entanglement}, collectively provide theoretical foundations that guided parameter selection in our simulations. We extended the ideas of these papers into the space of modular quantum systems, demonstrating potential yield and performance benefits that MCMs offer.

Application-specific architecture design for SC QCs was explored in~\cite{li2020towards} with a bottom-up approach. This prior work contributes a methodology that analyzes benchmarks for connectivity and optimizes for communication and qubit frequency allocation when mapping to transmon QCs. Resulting QC architectures of 20 qubits or less were analyzed in terms of both yield and mapped algorithm gate counts. While~\cite{li2020towards} motivates researching the balance between QC performance and hardware yield, our work differs by taking a more top-down approach to explore the practicality of modular quantum systems, quantum MCMs, as QCs scale to hundreds of qubits. We develop a MCM simulation tool, currently parameterized with state-of-art fabrication and real machine details, but adaptable to future QC improvements. We target a wide range of architecture sizes during yield, device property, and performance analysis.

\section{Conclusion}
\label{8-conclusion}

Quantum machines have rapidly improved in fidelity and qubit count, but they face significant scaling challenges.  Here, we propose a modular approach 
as a strategy to overcome the obstacles associated with fabricating larger QCs.  Our work presents a first step toward modeling the  yield, fidelity, and performance properties of fixed-frequency, transmon-based quantum chiplet architectures, exploring tradeoffs in system design.  
Our results show that chiplet architectures, relative to monolithic designs, benefit from average yield improvements ranging from  approximately $ 9.6 - 92.6\times$ for $\lesssim$500 qubit machines. In addition, we discover MCM configurations that demonstrate average two-qubit gate infidelity reductions of approximately $ 0.949- 0.815\times$ compared to monolithic counterparts. We hope that this work leads to greater exploration of quantum chiplet design.

\section*{Acknowledgment}

The authors would like to thank Malcolm Carroll, Nathan Earnest, Nikos Hardavellas, Jared Hertzberg, Ali Javadi-Abhari, and Jason Orcutt for their helpful discussions and comments during the preparation of this manuscript. This work is funded in part by EPiQC, an NSF Expedition
in Computing, under award CCF-1730449; in part
by STAQ under award NSF Phy-1818914; in part by NSF
award 2110860; in part by the US Department of Energy Office 
of Advanced Scientific Computing Research, Accelerated 
Research for Quantum Computing Program; in part by the 
NSF Quantum Leap Challenge Institute for Hybrid Quantum Architectures and 
Networks (NSF Award 2016136); and in part based upon work supported by the 
U.S. Department of Energy, Office of Science, National Quantum 
Information Science Research Centers.  This research used resources of the Oak Ridge Leadership Computing Facility, which is a DOE Office of Science User Facility supported under Contract DE-AC05-00OR22725. KNS is supported by IBM as a Postdoctoral Scholar at the University of Chicago and the Chicago Quantum Exchange. GSR is supported as a Computing Innovation Fellow at the University of Chicago. This material is based upon work supported by the National Science Foundation under Grant \# 2030859 to the Computing Research Association for the CIFellows Project. FTC is Chief Scientist for Quantum Software at ColdQuanta and an advisor to Quantum Circuits, Inc.

\bibliographystyle{IEEEtranS}
\bibliography{refs}



%

\end{document}